\documentclass[twocolumn,amsmath,aps]{revtex4}

\usepackage{amssymb,mathrsfs,bm,color,times}
\usepackage{graphicx,color}
\usepackage{bm}
\usepackage[hypertex]{hyperref}
\usepackage{amsmath}

\newcommand{\be}{\begin{equation}}
\newcommand{\ee}{\end{equation}}
\newcommand{\bea}{\begin{eqnarray}}
\newcommand{\eea}{\end{eqnarray}}
\newcommand{\bsube}{\begin{subequations}}
\newcommand{\esube}{\end{subequations}}

\newcommand{\Eq}[1]{Eq.\,(\ref{#1})}

\newcommand{\la}{\langle}
\newcommand{\ra}{\rangle}

\newcommand{\beq}{\begin{equation}}
\newcommand{\eeq}{\end{equation}}
\newcommand{\beqn}{\begin{eqnarray}}
\newcommand{\eeqn}{\end{eqnarray}}
\newcommand{\bsub}{\begin{subequations}}
\newcommand{\esub}{\end{subequations}}

\begin{document}
%\begin{CJK*}{GBK}{Song}

\title{Enhanced super-Heisenberg scaling precision
by nonlinear coupling and postselection }

\author{Lupei Qin }
\email{qinlupei@tju.edu.cn}
\affiliation{Center for Joint Quantum Studies and Department of Physics,
School of Science, \\ Tianjin University, Tianjin 300072, China}

\author{Jialin Li}
\affiliation{Center for Joint Quantum Studies and Department of Physics,
School of Science, \\ Tianjin University, Tianjin 300072, China}

\author{Yazhi Niu}
\affiliation{Center for Joint Quantum Studies and Department of Physics,
School of Science, \\ Tianjin University, Tianjin 300072, China}

\author{Xin-Qi Li}
\email{xinqi.li@tju.edu.cn}
\affiliation{Center for Joint Quantum Studies and Department of Physics,
School of Science, \\ Tianjin University, Tianjin 300072, China}

\date{\today}

%% \maketitle
\begin{abstract}
In quantum precision metrology, the famous result of Heisenberg limit
scaling as $1/N$ (with $N$ the number of probes)
can be surpassed by considering nonlinear coupling measurement.
In this work, we consider the most practice-relevant quadratic nonlinear coupling
and show that the metrological precision can be enhanced
from the $1/N^{\frac{3}{2}}$ super-Heisenberg scaling to $1/N^2$,
by simply employing a pre- and post-selection (PPS) technique,
but not using any expensive quantum resources such as quantum entangled state of probes.
\end{abstract}

% \pacs{03.65.Yz,03.65.Sq,31.15.xv,31.15.xg}

\maketitle

{\flushleft\it Introduction}.---
Quantum precision measurement (or, quantum metrology)
is one of the main quantum technology frontiers under explorations.
It can reach unprecedented metrological precision
owing to using quantum resources such as entanglement and squeezing \cite{Ld04,Ld06,Ld11}.
However, the precision of quantum metrology is also limited by
measurement strategies and some fundamental quantum principles.
For instance, the quantum nature of the underlying state, interacting evolution,
and final measurement would make the output results of measurement
suffer strong statistical uncertainty.
Fundamentally, the various metrological precisions
must be bounded by the Heisenbergy uncertainty principle
or, equivalently, the Heisenberg inequality \cite{Lui04}.
%% ===========
For the metrological schemes taking only classical strategies,
i.e., not exploring any quantum correlations between the probes,
the precision can at most scale with $N$ (the number of probes)
as $1/\sqrt{N}$ \cite{Ld04,Ld06,Ld11,Lui04}.
This is actually the shot-noise-caused scaling limit
(owing to the absence of correlation),
which is also, more often, referred to as standard quantum limit (SQL),
owing to taking standard classical measurement strategies.
%%  ================
However, the standard classical strategies are not optimal.
If quantum correlation is introduced, e.g., by quantum entanglement of the probes,
the SQL can be surpassed by achieving the better precision of $1/N$ scaling,
which has a $\sqrt{N}$ precision enhancement over the SQL \cite{Ld04,Ld06,Ld11}.
This scaling is usually referred to as Heisenberg limit (HL).
For long time, HL was believed as the ultimate limit,
e.g., as the ultimate bound to the precision of phase measurements
in the case of interferometry.

However, the above conclusions of SQL and HL are achieved only for linear coupling measurement.
The so-called HL (scaling as $1/N$, but not really limited by the Heisenberg principle)
can be surpassed by considering nonlinear coupling measurements
\cite{Lui04,Lui06,Lui07,Cav07,Cav08a,Cav08b,Cav09}.
In this work, following some references,
we may dub the metrological precision better than $1/N$ scaling
as super-Heisenberg limit (or super-Heisenberg scaling).
For a general $k_{\rm th}$-order nonlinear coupling measurement,
the main conclusions are \cite{Lui04,Lui06,Lui07,Cav07,Cav08a,Cav08b,Cav09}:
{\it (i)}
an optimal sensitivity that scales as $1/N^k$ can be achieved,
if using as well an entangled initial probe state;
{\it (ii)}
sensitivity that scales as $1/N^{k-1/2}$ can be possible,
if the probe is initially prepared in a product state.
In practice, nonlinear coupling can be realized in such as
quantum optical and condensed matter systems.
For instance, following the pioneering works,
subsequent studies consider the specific realizations
by introducing nonlinear Kerr medium in the Mach-Zehnder interferometer (MZI),
and by exploring spin-based atom ensembles and condensed matter systems
\cite{Lui15,Zhao19,Shaj21,Mit10}.
Moreover, experimental demonstrations have
also been carried out \cite{Mit11,Mit14,Peng18,Yuan20}.
For nonlinear coupling measurement,
the most practice-relevant case is the second-order nonlinearity with $k=2$.
The both results scaling as $1/N^2$ (for entangled initial probe state)
and $1/N^{\frac{3}{2}}$ (for product initial probe state)
are super-Heisenberg scaling, being thus of great interest.

In this work, we consider to introduce the strategy of pre- and post-selection (PPS),
which was proposed by Aharonov, Albert, and Vaidman (AAV)
in the quantum weak value (WV) measurement \cite{AAV1,AAV2}.
We will show that, without need of additional quantum resources
such as entangled state of the probes,
the PPS strategy can help to enhance the metrological precision scaling
from $1/N^{\frac{3}{2}}$ to $1/N^2$, for the quadratic nonlinear coupling measurement.
Actually, the weak-value-amplification (WVA) technique has been successfully
demonstrated in quantum precision measurement in various contexts
\cite{Kwi08,How09a,How10a,How10b,How13,Sto12,Lun17,ZLJ20}.
The basic conclusion achieved by the WVA community is somehow twofold.
In one aspect, indeed, the WVA technique
can amplify small signals beyond the resolution of detector,
can suppress the technical noise in some practical situations,
and can greatly outperform the conventional measurement (CM)
in the presence of power saturation of detectors
\cite{Sim10,Ste11,Nish12,Bru15,Bru16,How17,Jor17,Pan18,How15,Jor18}.
In other aspect, owing to discarding data in the postselection,
the Fisher information encoded in the postselected data
(and thus the metrological precision)
cannot surpass the result of conventional measurement
(without involving the postselection procedure) \cite{Tana13,FC14,Kne14,Ked12,Jor14,Ren20}.

Very recently, by considering a two-state system (qubit)
linearly coupled to an optical coherent state (probe meter),
it was found that the PPS strategy employed in the WVA measurement
can lead to some new results \cite{XQLi22}:
(i) with the increase of the measurement strength (to violate the AAV limit),
the WVA scheme can outperform the conventional approach;
(ii) the WVA scheme can make a mixture of coherent states work better
than a pure coherent state with identical average photon numbers,
while the opposite conclusion was claimed
in context of conventional measurement (without using the PPS strategy) \cite{Lui10}.
In present work, we extend the study of Ref.\ \cite{XQLi22}
from linear to quadratic nonlinear coupling measurement,
and obtain new results over the conventional measurement,
by means of the PPS strategy of WVA.

%\clearpage
\vspace{0.1cm}
{\flushleft\it Formulation of the measurement scheme}.---
%%%%
Following Ref.\ \cite{XQLi22}, let us consider a quantum two-state (qubit) system,
with states denoted as $|1\rangle$ and $|2\rangle$,
coupled to an optical probe (meter) system prepared in the coherent state $|\alpha\rangle$.
However, in  this work, rather than the linear coupling,
we consider a quadratic nonlinear coupling described by
$\hat{H}^{'}=-\lambda\hat{\sigma}_{z}\hat{n}^2$,
where the Pauli operator
$\hat{\sigma}_{z}=|2\rangle\langle2|-|1\rangle\langle1|$ describes the qubit system,
and the number operator $\hat{n}=\hat{a}^{\dagger}\hat{a}$ is for the optical meter,
with $\hat{a}^{\dagger}$ ($\hat{a}$) the creation (annihilation) operator of a photon.
Accordingly, the evolution of the whole qubit-plus-meter system
is governed by the unitary operator
$U=\exp(i\chi \hat{\sigma}_z\hat{n}^2)$,
where $\chi=\int_0^\tau dt \lambda= \lambda \tau$
is the integrated strength of interaction over time $\tau$.
For quantum precision measurement, we assume $\chi$ the parameter to be estimated.
Starting with a product state of the qubit and meter,
$|\Psi^{(i)}_J\rangle=|i\rangle|\alpha\rangle$,
where $|i\rangle=\cos\frac{\theta_i}{2}
|1\rangle+\sin\frac{\theta_i}{2} e^{i\varphi_i}|2\rangle$,
the coupling interaction will evolve the entire system into an entangled state as
\begin{equation}\label{}
 |\Psi_J\rangle=\cos\frac{\theta_i}{2}
 |1\rangle|\phi_{-}\rangle
 +\sin\frac{\theta_i}{2}e^{i\varphi_i}
 |2\rangle|\phi_+\rangle  \,,
\end{equation}
where the meter's state is affected differently
by the qubit states $|1\ra$ and $|2\ra$, being given by
\begin{equation}\label{phi-+}
|\phi_{\mp}\rangle=e^{-|\alpha|^2/2}
\sum_{n=0}^{\infty}\frac{\alpha^{n}}{\sqrt{n!}}
e^{\mp i\chi n^2}|n\rangle  \,.
\end{equation}
Then, through the coupling interaction, the parameter $\chi$ under estimation
has been encoded into the meter states.
One can perform various types of measurement to extract the value of $\chi$.
In this work, following the idea of WVA, we consider
applying the strategy of post-selection to amplify the signal for $\chi$.
To be specific, let us assume the post-selection measurement of the qubit state with
$|f\rangle=\cos\frac{\theta_f}{2}|1\rangle +\sin\frac{\theta_f}{2}e^{i\varphi_f}|2\rangle$,
and keep the final measurement results (e.g., photon numbers) of the meter
only after successful post-selection of the qubit state.
Mathematically, the post-selection is simply described as
$|\widetilde{\Phi}_f\rangle=\langle f|\Psi_J\rangle$,
resulting thus in a state for the meter as
\begin{equation} \label{pstate}
 |\widetilde{\Phi}_f\rangle
 =\cos\frac{\theta_i}{2}
 \cos\frac{\theta_f}{2}|\phi_{-}\rangle
 +\sin\frac{\theta_i}{2}
 \sin\frac{\theta_f}{2}e^{i\varphi_0}|\phi_+\rangle  \,.
\end{equation}
Here $\varphi_0$ is introduced to denote the phase difference
of the PPS states $|i\ra$ and $|f\ra$, i.e., $\varphi_0=\varphi_i-\varphi_f$.
We further normalize the postselected meter state as
$|\Phi_f\rangle=|\widetilde{\Phi}_f\rangle/\sqrt{p_f}$,
with $p_f=\langle \widetilde{\Phi}_f|\widetilde{\Phi}_f\rangle$.
We may mention that $p_f$ is right the success probability of postselection,
which is explicitly given by
\begin{equation}
\label{postp}
  p_f=A+B e^{-|\alpha|^2} \sum_{n=0}^{\infty}\frac{|\alpha|^{2n}}{n!}
  \cos(2\,\chi\,n^2+\varphi_0) \,,
\end{equation}
where $A=\frac{1}{2}(1+\cos{\theta_i}\cos{\theta_f})$
and $B=\frac{1}{2}\sin{\theta_i}\sin{\theta_f}$.
Based on the state $|\Phi_f\rangle$,
one can consider certain specific measurement,
e.g., the photon number measurement to it.
Straightforwardly, the probability of obtaining $n$ photons
is given by $P_f(n)=|\langle n|\Phi_f\rangle|^2$.
Explicitly, we find
\begin{equation}\label{}
  P_f(n)=\frac{e^{-|\alpha|^2}}{p_f}
  \left(\frac{|\alpha|^{2n}}{n!}\left[A+B \cos(2\,\chi\,n^2+\varphi_0)\right]\right) \,.
\end{equation}
In order to carry out a characterization for the metrological precision of the parameter $\chi$,
we can calculate the Fisher information (FI) $F_f$ of $\chi$
encoded in the distribution function $P_f(n)$, through
\begin{equation}\label{}
  F_f=\sum_{n}\frac{1}{P_f(n)}\left(\frac{\partial P_f(n)}{\partial \chi}\right)^2\,.
\end{equation}
Then, the estimate precision for $\chi$
is bounded as $\delta(\chi)\ge 1/ \sqrt{p_f F_f}$,
following the Cram\'{e}r-Rao bound (CRB) inequality.
This is the result associated with the specific scheme
of photon-number measurement to $|\Phi_f\rangle$.
One can also consider the so-called quantum Fisher information (QFI)
encoded in the state $|\Phi_f\rangle$, which is given by
\begin{equation}
\label{QFI}
  Q_f=4\left(\frac{d \langle \Phi_f|}{d \chi}\frac{d |\Phi_f\rangle}{d \chi}
  -\left|\langle \Phi_f|\frac{d |\Phi_f\rangle}{d \chi}\right|^2\right) \,.
\end{equation}
Unlike the FI associated with certain specific measurement scheme,
the QFI represents the maximum amount of information
for all possible measurement schemes.
Thus, we carry out the following result for the WVA scheme
\begin{align} \label{WVA-QFI}
p_fQ_f&=4\,\left[A\langle \hat{n}^4\rangle -\frac{1}{p_f}
\left(C \langle\hat{n}^2 \rangle\right)^2 \right.\\ \nonumber
  &-B e^{-|\alpha|^2} \sum_{n=0}^{\infty}
   \frac{|\alpha|^{2n}}{n!}\,n^4\,\cos(2\,\chi\,n^2+\varphi_0) \\ \nonumber
  &-\left. \frac{1}{p_f} B^2 \left(e^{-|\alpha|^2} \sum_{n=0}^{\infty}
   \frac{|\alpha|^{2n}}{n!}\,n^2\,\sin(2\,\chi\,n^2+\varphi_0)\right)^2
    \right]  \,.
\end{align}
In this result, $\langle \bullet\rangle$
represents the expectation over the optical coherent state $|\alpha\rangle$.
Thus we have $\langle \hat{n}^4\rangle=N^4+6N^3+4N^2+N$,
and $\langle \hat{n}^2\rangle=N^2+N$.
Here we also introduced $C=(\cos{\theta_i}+\cos{\theta_f})/2$.

In this work, we would like to term the FI (QFI) $p_fF_f$ ($p_fQ_f$) as WVA-FI (WVA-QFI),
and especially in numerical demonstrations choose the PPS parameters as
$\theta_i=\frac{\pi}{2}$, $\varphi_0=\pi$, and $\theta_f=\frac{\pi}{2}$.
This choice makes the WVA in the AAV regime, i.e., making the
AAV-WV $\sigma_z^w=\la f|\sigma_z|i\ra / \la f|i\ra$ divergent.
Note that, for finite coupling strength, the postselected average of readout results
is not divergent, but maximally amplified \cite{Ren20}.
In the AAV limit, one can prove that the meter's state after postselection is
$\widetilde{\Phi}_f \sim \la f|i\ra e^{i\chi\sigma_z^w \hat{n}^2}|\alpha\ra$,
from which we find that the coupling strength is amplified as
$\widetilde{\chi}=\chi\sigma_z^w$. Actually, this reveals the basic idea of WVA.

In order to carry out a comparison with the conventional measurement,
let us consider to encode the parameter $\chi$ into the probe state $|\alpha\ra$ by coupling
through a single state $|1\ra$ or $|2\ra$ of the qubit, but not through their superposition.
Actually, in the absence of postselection,
it can be proved that using the superposed state of $|1\ra$ and $|2\ra$
will result in the same QFI as using the single state $|1\ra$ or $|2\ra$ \cite{XQLi22}.
To be specific, let us consider coupling the qubit through $|1\rangle$ to the probe field.
In this case, after interaction, the joint state for the entire qubit-plus-meter is
$|\Psi\rangle_{\rm{cm}}=|1\rangle|\phi_-\rangle$.
Then, the QFI of $\chi$ in the state $|\phi_-\rangle$ is simply obtained as
\begin{equation}\label{CM-QFI}
  Q_{\rm{cm}}=4\left(4\,N^3+6\,N^2+N\right) \,.
\end{equation}
In this result, we also use $N=|\alpha|^2$,
to denote the mean photon number of the probe field.

\begin{figure}
  \centering
  \includegraphics[scale=0.3]{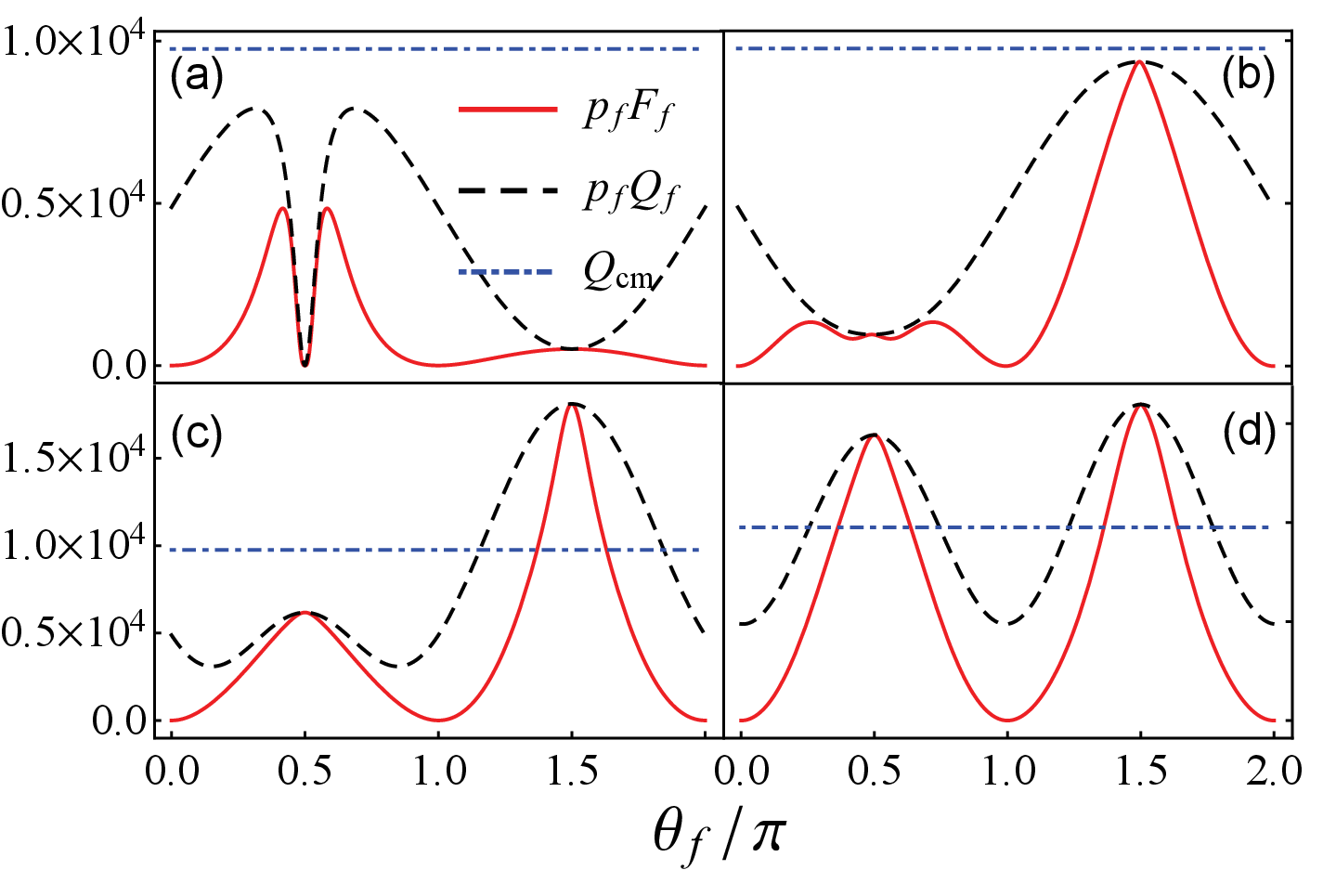}\\
  \caption{
The WVA-FI $p_f F_f$ (solid red lines) and WVA-QFI $p_f Q_f$ (dashed black lines)
are compared with each other, and with the QFI $Q_{\rm cm}$ of conventional measurement
(dash-dotted blue lines, without using the PPS technique).
The results shown in (a)-(d) correspond to measurement strengths
$\chi=0.001$, 0.005, 0.01, and 0.1, respectively.
Other parameters used for the numerical calculations are:
$\theta_i=\frac{\pi}{2}$, $\varphi_0=\pi$, and $|\alpha|^2=N=8$.   }
\label{fig1}
\end{figure}

\vspace{0.1cm}
{\flushleft\it Results and discussion}.---
%%%%
For linear coupling measurement, as studied in Ref.\ \cite{XQLi22},
the interaction-strength-encoded meter states $|\phi_{\pm}\ra$ have simple analytic solutions.
However, for the quadratic nonlinear coupling measurement considered in this work,
we are unable to find analytic solutions for $|\phi_{\pm}\ra$, but only present
the formal expression in terms of power series expansion, as shown by \Eq{phi-+}.
Therefore, we would like to carry out numerical results and make some discussions.

In Fig.\ 1 we plot the numerical results of $p_f F_f$ and $p_f Q_f$ for the WVA measurement.
Indeed, as expected, we find that $p_f F_f$ is always bounded by $p_f Q_f$.
Only at some special postselection angles, e.g., at $\theta_f=3\pi/2$,
$p_f F_f$ can reach the maximum value of $p_f Q_f$.
Since the WVA-FI and WVA-QFI are periodic functions of $\theta_f$ with period of $2\pi$,
here we only show the results of a single period.
We notice that the postselection dependence behavior
varies drastically with the measurement strength $\chi$.
In Fig.\ 2 we show the particular $\chi$ dependence
which, non-monotonically oscillating in the stronger coupling regime,
is quite different from the linear coupling measurement
(see Fig.\ 2 in Ref.\ \cite{XQLi22}).

Importantly, we find that, with the increase of $\chi$,
both $p_fQ_f$ and $p_fF_f$ can exceed the QFI $Q_{\rm cm}$ of the conventional scheme,
which is free from $\chi$, as observed in Fig.\ 1 (c) and (d).
This result is somehow beyond the usual claim among the WVA community
\cite{Nish12,Ked12,Jor14,Ren20,Tana13,FC14,Kne14}.
That is, it was generally believed that,
from either the perspective of Fisher information or the signal-to-noise ratio,
the {\it intrinsic} overall estimate precision cannot be enhanced by the WVA technique.
The basic reason is that, despite the enhanced signal after postselection,
the postselection will discard a large number of measurement data
and thus result in a larger uncertainty fluctuations.
Therefore, the WVA technique was conceived of having, at least,
no theoretical advantages \cite{Tana13,FC14,Kne14},
but only having some technical advantages in practice such as
not being bounded by the saturation limit of photo-detectors \cite{Lun17,ZLJ20}.

We notice that the analysis of achieving such conclusion was largely based on
using the transverse spatial wavefunction of a light beam as probe meter,
for instance, in Ref.\ \cite{Jor14}, and in many other references.
In Ref.\ \cite{XQLi22} and in present work,
the analysis is fully along the line in Ref.\ \cite{Jor14}.
The only difference is using an optical coherent state as probe meter.
Therefore, the claim that the WVA scheme cannot exceed the conventional measurement
is not a result imposed by fundamental physics.
The result displayed in Fig.\ 1 indicates that the nonlinear coupling WVA measurement
can reach a precision better than the conventional approach
in the absence of postselection.
This is a next example, subsequent to the previous
investigation of linear coupling measurement \cite{XQLi22}.

We also notice that the conventional nonlinear coupling measurement (without postselection)
can realize super-Heisenberg scaling metrological precision.
For instance, for quadratic nonlinear coupling measurement, it was found
\cite{Lui04,Lui06,Lui07,Cav07,Cav08a,Cav08b,Cav09}
that the precision can scale with the (average) photon numbers of probe
as $1/N^{\frac{3}{2}}$.
Indeed, this is in agreement with the result of QFI shown in \Eq{CM-QFI},
which scales with the photon numbers dominantly as $N^3$ in the large $N$ limit.
In this context, we may raise an interesting and practically important question:
Is it possible to surpass this super-Heisenberg scaling limit,
under the restriction of quadratic nonlinear coupling measurement
and not using expensive quantum resources such as entangled state of the probe photons?
In the following, we will show that the answer is {\it yes}.
That is, by involving the simple PPS strategy in WVA,
we can boost the metrological precision
to scale with the probe photon numbers from $1/N^{\frac{3}{2}}$ to $1/N^2$.
Actually, preliminary insight from the result of the WVA-QFI given by \Eq{WVA-QFI}
indicates that this is possible,
owing to the leading term proportional to $N^4$ in large $N$ limit.
The only unclear point is how fast this term becomes the dominant one
after accounting for the summed pre-coefficients, when compared with
other lower-order terms in the relatively complicated expression of \Eq{WVA-QFI}.

In Fig.\ 1, we have already shown that
the WVA-FI can exceed the QFI of the conventional scheme
for a fixed average photon number $N$ of the probe field,
by properly (even slightly) increasing the measurement coupling strength $\chi$.
Now, we add to show in Fig.\ 3 that for a fixed coupling strength $\chi$,
the WVA-FI $p_f F_f$ can increase very fast with the average probe photon numbers.
Actually, the increasing is highly nonlinear, owing to the nonlinear measurement coupling.
This implies that the precision enhancement
boosted by the PPS technique of WVA
will be very efficient for the nonlinear coupling measurement,
by increasing the photon numbers of the probe field. 
We also notice that the WVA-FI enhancement with the increase of $N$
is more efficient for stronger nonlinear coupling strength
(but being restricted as well in the weak coupling regime).
This can be understood jointly with the help of the result in Fig.\ 2.

\begin{figure}
  \centering
  \includegraphics[scale=0.7]{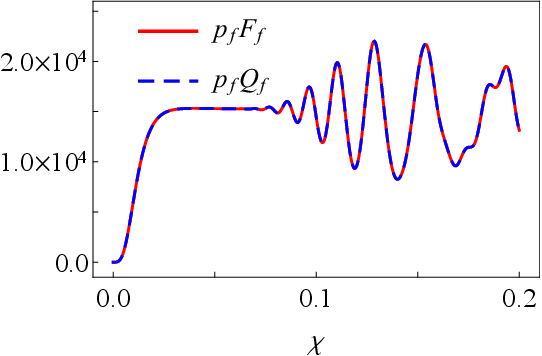}\\
  \caption{
Measurement strength dependence of the WVA-FI $p_fF_f$ (solid red line)
and WVA-QFI $p_fQ_f$ (dashed blue line).
The WVA related PPS parameters are chosen as
$\theta_i=\frac{\pi}{2}$, $\varphi_0=\pi$, and $\theta_f=\frac{\pi}{2}$;
and the average photon number is assumed as $|\alpha|^2=N=8$.  }
\label{fig2}
\end{figure}

\begin{figure}
  \centering
  \includegraphics[scale=0.75]{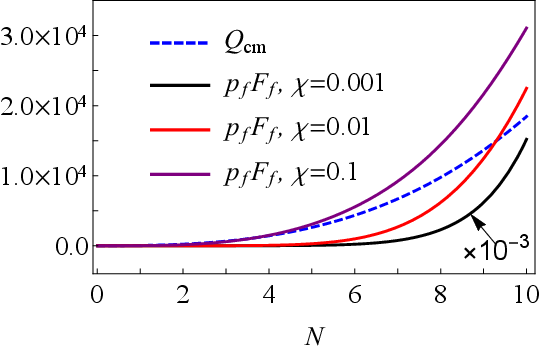}\\
  \caption{
Average photon number dependence of the WVA-FI $p_fF_f$
and the conventional QFI $Q_{{\rm cm}}$.
Results shown are for measurement strengths $\chi=0.001$, 0.01, and 0.1,
while the result of $\chi=0.001$ in the plot has been amplified by $10^{3}$
for better visual effect.
Note also that the conventional QFI $Q_{{\rm cm}}$ is independent of $\chi$.
The WVA related PPS parameters are the same as in Fig.\ 2.   }
\end{figure}

\begin{figure}
  \centering
  \includegraphics[scale=0.7]{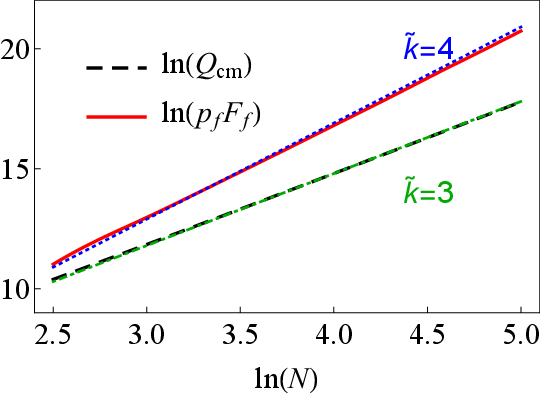}\\
  \caption{
Super-Heisenberg scaling with the average photon number,
re-plotted using the result of measurement strength $\chi=0.01$
in Fig.\ 3 (but for larger range of $N$).
The numerical results are fitted by the solid straight lines
with slopes $\widetilde{k}=4$ and $3$, respectively.
The slope fitted here corresponds to the precision scaling $1/N^k$
with $k=\widetilde{k}/2$, according to the CRB inequality. 
Then, it is evident that the WVA-FI $p_fF_f$ has better precision scaling
than the conventional QFI $Q_{{\rm cm}}$,
i.e., with a $\sqrt{N}$ precision enhancement
purely owing to employing the PPS strategy in WVA.     }
\end{figure}

In Fig.\ 4, we re-plot the result of the measurement strength $\chi=0.01$ in Fig.\ 3
to show super-Heisenberg scaling with the average photon numbers
(for larger range of $N$, but not much large, for instance, approximately larger than ten).
We explicitly demonstrate the dominant
overall $N^4$-scaling behavior for the WVA-FI $p_f F_f$,
in comparison with the $N^3$-scaling of the QFI $Q_{\rm cm}$ of the conventional scheme.
The WVA-FI scaling behavior shown in Fig.\ 4 simply means that
the metrological precision can be enhanced by the PPS strategy of WVA
from $1/N^{\frac{3}{2}}$ to $1/N^2$ scaling limit, based on the CRB inequality.
% ====================
In this context, we may mention the long standing issue
of achieving the HL (beyond the SQL) in quantum metrology,
which is extremely demanding, e.g.,
needing expensive resources such as quantum entanglement or squeezing.
Noting that what has been gained from SQL to HL is a $\sqrt{N}$ precision enhancement,
it seems thus extremely desirable to achieve this same precision enhancement
by means of the simple PPS strategy of WVA.

About the quadratic nonlinear coupling, it can be realized by various ways
based on available state-of-the-art experimental platforms.
For instance, in the Mach-Zehnder interferometer (MZI),
inserting nonlinear optical kerr medium into the interference arms
can lead to quadratic nonlinear coupling with the probe laser field,
while the laser field is well described by a coherent state.
Since the interference paths can be described as the `system' states $|1\ra$ and $|2\ra$,
the nonlinear MZI can fall into the theoretical description of our present work.
Recently, based on the nonlinear MZI, interesting studies
were carried out for parameter estimation
of the gravitational effect, such as estimating
the characteristic parameters of black-hole and wormholes \cite{Ral17,Sab17,Sab18}.
In particular, the super-Heisenberg scaling $\sim 1/N^{\frac{3}{2}}$
was highlighted \cite{Ral17,Sab17,Sab18},
viewing that current techniques (e.g. the atomic interferometry)
for gravitational measurements are limited to the SQL.
Even if using quantum entanglement or squeezing resource,
only the HL precision can be achieved.
Our present study, remarkably, indicates that,
if further applying the PPS strategy of WVA,
the performance of the nonlinear MZI can be improved
to achieve the super-Heisenberg scaling $\sim 1/N^2$.

In addition to the realization in terms of the nonlinear optical MZI mentioned above,
it is also possible to realize the quadratic nonlinear coupling
as considered in this work
by employing the optical cavity QED or solid-state circuit QED systems.
For instance, consider a three-level atom (or artificial atom)
coupled with the single-mode cavity field.
For the three-level atom, beside the ground state $|g\ra$
and excited state $|e\ra$,
we assume the existence of
a third intermediate state $|e'\ra$
for the purpose of mediating a two-photon
virtual transition coupling between $|g\ra$ and $|e\ra$,
which can result thus in the nonlinear dispersive coupling Hamiltonian
$H'=\chi \sigma_z(a^{\dagger} a)^2$,
where $\sigma_z=|e\ra\la e|-|g\ra\la g|$.
The nonlinear coupling strength $\chi$ reads as
$\chi=|g_1 g_2|^2/(\Delta_1^2 \Delta_2)$,
obtained by using the 4th-order perturbation theory.
In this result, $g_1$ and $g_2$ are the single photon transition coupling strengths
between the states $|g\ra$ and $|e'\ra$,
and between $|e'\ra$ and $|e\ra$, respectively.
The energy detunings are $\Delta_1=\omega-(E_{e'}-E_g)$
and $\Delta_2=2\omega-(E_e-E_g)$.
We also mention that the interplay of cavity driving and photon loss
will result in a coherent state in steady state.
Then in the so-called `bad-cavity' limit,
the coupling of the cavity field to the atom
and the subsequent measurement of the field \cite{Kor11,Kor16,Li15,Li16},
are completely captured by the treatment in present work.\\
\\
{\it Summary}.---
%%%%
For the purpose of quantum metrology,
we have shown that, without need of expensive quantum resources
such as entanglement of probes or squeezing,
applying the PPS strategy of WVA
can help to enhance the metrological precision scaling with the probe photon numbers
from $1/N^{\frac{3}{2}}$ to $1/N^2$, for quadratic nonlinear coupling measurement.
The basic reason of achieving this enhancement is that
the PPS technique of WVA encodes the parameter $\chi$
into the superposed state of \Eq{pstate},
resulting thus in the dominant $N$ scaling of QFI from $N^3$ to $N^4$,
which has a $\sqrt{N}$ precision enhancement over the conventional approach.
This better scaling means that
the metrological precision can be enhanced more efficiently
by increasing the photon numbers of the probe field,
while the overall WVA-FI is also larger than the QFI of conventional approach.
The metrology scheme analyzed in this work also enjoys
the general technical advantages of WVA,
such as breaking through the resolution limit of detector,
owing to the amplified `signal' (from $\chi$ to $\widetilde{\chi}$).

The nonlinear metrology scheme is quite relevant to phase measurement
based on the the Mach-Zehnder interferometer (MZI),
by considering to insert a piece of
nonlinear optical kerr medium into one of the interference arms,
which also resembles the laser interferometer
for gravitational wave detection, such as LIGO.
Thus, LIGO may consider the use of nonlinear media
and may even consider to employ the PPS strategy of WVA.
The PPS procedure of WVA in the MZI can be implemented by simply
introducing a controllable phase shift in one of the arms,
before collecting the probe light from one output port
to perform photon number (intensity) or quadrature measurement.
We anticipate that the present work
can motivate further exploration along this research line.

% ============================================
\vspace{0.5cm}
{\flushleft\it Acknowledgements.}---
This work was supported by
the NNSF of China (Nos.\ 11675016, 11974011 \& 61905174).

\vspace{0.5cm}
% \clearpage
% =====================================================

%\end{CJK*}

\begin{references}



%%(00)
%%  {Ld04,Ld06,Ld11}
\bibitem{Ld04}
V. Giovannetti, S. Lloyd, and L. Maccone,
{\it Quantum-enhanced measurements: Beating the standard quantum limit},
Science \textbf{306}, 1330(2004).

\bibitem{Ld06}
V. Giovannetti, S. Lloyd, and L. Maccone,
{\it Quantum metrology},
Phys. Rev. Lett. \textbf{96}, 010401 (2006).

\bibitem{Ld11}
V. Giovannetti, S. Lloyd, and L, Maccone,
{\it Advances in quantum metrology},
Nat. Photon. \textbf{5}, 222 (2011).
%
%(I)
%早期 pioneering  works .. nL
%==================================
\bibitem{Lui04}  %{Lui04,Lui06,Lui07,Cav07,Cav08a,Cav08b,Cav09}
A. Luis,
{\it Nonlinear transformations and the Heisenberg limit},
Phys. Lett. A \textbf{329}, 8 (2004).

\bibitem{Lui06}
J. Beltr{\' a}n and A. Luis,
{\it Breaking the Heisenberg limit with inefficient detectors},
Phys. Rev. A \textbf{72}, 045801 (2005).

\bibitem{Lui07}
A. Luis,
{\it Quantum limits, nonseparable transformations, and nonlinear optics},
Phys. Rev. A \textbf{76}, 035801 (2007).

\bibitem{Cav07}
S. Boixo, S. T. Flammia, C. M. Caves, and J. M. Geremia,
{\it Generalized limits for single-parameter quantum estimation},
Phys. Rev. Lett. \textbf{98}, 090401 (2007).

\bibitem{Cav08a}
S. Boixo, A. Datta, M. J. Davis, S. T. Flammia, A. Shaji, and C. M. Caves,
{\it Quantum metrology: Dynamics versus entanglement},
Phys. Rev. Lett. \textbf{101}, 040403 (2008).

\bibitem{Cav08b}
S. Boixo, A. Datta, S. T. Flammia, A. Shaji, E. Bagan, and C. M. Caves,
{\it Quantum-limited metrology with product states},
Phys. Rev. A \textbf{77}, 012317 (2008).

\bibitem{Cav09}
S. Boixo, A. Datta, M. J. Davis, A. Shaji, A. B. Tacla, and C. M. Caves,
{\it Quantum-limited metrology and Bose-Einstein condensates},
Phys. Rev. A \textbf{80}, 032103 (2009).
%(II)
%later studies focus on:  nL-MI/MZI (引入Kerr 介质)
%           +  and spin-based system: nonlinear spin-spin interaction ...
%===========================================================================
\bibitem{Lui15}    % {Lui15,Zhao19,Shaj21,Mit10,Deng21}
A. Luis, and {\' A}. Rivas,
{\it Nonlinear Michelson interferometer for improved quantum metrology},
Phys. Rev. A \textbf{92}, 022104 (2015).

\bibitem{Zhao19}
J.-D. Zhang, Z.-J. Zhang, L.-Z. Cen, J.-Y. Hu, and Y. Zhao,
{\it Nonlinear phase estimation:
Parity measurement approaches the quantum Cram{\' e}r-Rao bound for coherent states},
Phys. Rev. A \textbf{99}, 022106 (2019).
%?? 此文 为实验
%3. G.-F. Jiao, K. Zhang, L. Q. Chen, W. Zhang, and C.-H. Yuan, Nonlinear phase estimation enhanced by an actively correlated Mach-Zehnder interferometer, Phys. Rev. A \textbf{102}, 033520 (2020)\\
\bibitem{Shaj21}
B. J. Mathew and A. Shaji,
{\it Nonclassical states of light in a nonlinear Michelson interferometer},
Phys. Rev. A \textbf{104}, 062604 (2021).

\bibitem{Mit10}
M. Napolitano and M. W. Mitchell,
{\it Nonlinear metrology with a quantum interface},
New J. Phys. \textbf{12}, 093016 (2010).

%\bibitem{Deng21}
%X. Deng, S.-L. Chen, M. Zhang, X.-F. Xu, J. Liu, Z. Gao, X.-C. Duan, M.-K. Zhou, L. Cao, Z.-K. Hu, {\it
%Quantum metrology with precision reaching beyond-$1/N$ scaling through $N$-probe
%entanglement generating interactions},
%Phys. Rev. A \textbf{104}, 012607 (2021).
%????  具体  啥内容 ???

%(III)
%nL .. super-HL  实验demonstration
%====================================
\bibitem{Mit11}  % {Mit11,Mit14,Peng18,Yuan20}
M. Napolitano, M. Koschorreck, B. Dubost, N. Behbood, R. J. Sewell, and M. W. Mitchell, {\it Interaction-based quantum metrology showing scaling beyond the Heisenberg limit}, Nature (London) \textbf{471}, 486 (2011).

\bibitem{Mit14}
R. J. Sewell, M. Napolitano, N. Behbood, G. Colangelo, F. M. Ciurana, and M. W. Mitchell,
{\it Ultrasensitive atomic spin measurements with a nonlinear interferometer},
Phys. Rev. X \textbf{4}, 021045 (2014).

\bibitem{Peng18}
X. Nie, J. Huang, Z. Li, W. Zheng, C. Lee, X. Peng, J. Du,
{\it Experimental demonstration of nonlinear quantum metrology with optimal quantum state}, Sci. Bull. \textbf{63}, 469 (2018).

\bibitem{Yuan20}
G.-F. Jiao, K. Zhang, L. Q. Chen, W. Zhang, and C.-H. Yuan,
{\it Nonlinear phase estimation enhanced by an actively correlated Mach-Zehnder interferometer},
Phys. Rev. A \textbf{102}, 033520 (2020).


\bibitem{AAV1}
Y. Aharonov, D. Z. Albert, and L. Vaidman,
{\it How the result of a measurement of a component of the spin of a spin-1/2 particle
can turn out to be 100},
Phys. Rev. Lett. {\bf 60}, 1351 (1988).

\bibitem{AAV2}
Y. Aharonov and L. Vaidman,
{\it Properties of a quantum system during the time interval between two measurements},
Phys. Rev. A {\bf 41}, 11 (1990).

%% =======================  WVA  实验 文献
%光的自旋霍尔效应
\bibitem{Kwi08} % {Kwi08}
O. Hosten and P. G. Kwiat,
{\it Observation of the Spin Hall Effect of Light via weak measurements},
Science {\bf 319}, 787 (2008).
% 测量微弱信号{How09a,How10a,How10b,How13,Sto12}
% How09a Ultrasensitive beam deflection
% How10a 相位放大
% How10b Precision frequency measurements
% How13  velocity measurements
% Sto12 temperature shift
\bibitem{How09a}
P. B. Dixon, D. J. Starling, A. N. Jordan, and J. C. Howell,
{\it Ultrasensitive beam deflection measurement via interferometric
weak value amplification},
Phys. Rev. Lett. {\bf 102}, 173601 (2009).
%\bibitem{How09b}
%D. J. Starling, P. B. Dixon, A. N. Jordan, and J. C. Howell,
%{\it Optimizing the signal-to-noise ratio of a beam-deflection
%measurement with interferometric weak values},
%Phys. Rev. A {\bf 80}, 041803 (2009).
\bibitem{How10a}
D. J. Starling, P. B. Dixon, N. S.Williams, A. N. Jordan, and J. C. Howell,
{\it Continuous phase amplification with a sagnac interferometer},
Phys. Rev. A {\bf 82}, 011802 (2010).

\bibitem{How10b}
D. J. Starling, P. B. Dixon, A. N. Jordan, and J. C. Howell,
{\it Precision frequency measurements with interferometric weak values},
Phys. Rev. A {\bf 82}, 063822 (2010).

\bibitem{How13}
G. I. Viza, J. Mart\'inez-Rinc\'on, G. A. Howland, H. Frostig,
I. Shomroni, B. Dayan, and J. C. Howell,
{\it Weak-values technique for velocity measurements},
Opt. Lett. {\bf 38}, 2949 (2013).

\bibitem{Sto12}
P. Egan and J. A. Stone,
{\it Weak-value thermostat with 0.2 mK precision},
Opt. Lett. {\bf 37}, 4991 (2012).

%==饱和情况下WVA优势 理论和实验 {Lun17,ZLJ20}
\bibitem{Lun17}
J. Harris, R. W. Boyd,and J. S. Lundeen,
{\it Weak value amplification can outperform conventional measurement in the presence of detector saturation},
Phys. Rev. Lett. {\bf 118} 070802 (2017).

\bibitem{ZLJ20}
L. Xu, Z. Liu, A. Datta, G. C. Knee, J. S. Lundeen, Y. Lu, and L. Zhang,
{\it Approaching quantum-limited metrology with imperfect detectors by using weak-value amplification},
Phys. Rev. Lett. {\bf 125}, 080501 (2020).

%%  \cite{Sim10,Ste11,Nish12,Ked12,Jor14,Bru15,Bru16,Jor17,How17,Ren20}.
% == {Sim10,Ste11,Nish12,Ked12,Jor14,Bru15,Bru16,Jor17,How17,Ren20} 信号放大、抵抗噪声
\bibitem{Sim10}
N. Brunner and C. Simon,
{\it Measuring small longitudinal phase shifts: weak measurements or standard interferometry?},
Phys. Rev. Lett. {\bf 105}, 010405 (2010).

\bibitem{Ste11}
A. Feizpour, X. Xingxing, and A. M. Steinberg,
{\it Amplifying single-photon nonlinearity using weak measurements},
Phys. Rev. Lett. {\bf 107}, 133603 (2011).

\bibitem{Nish12}
A. Nishizawa, K. Nakamura, and M. K. Fujimoto,
{\it Weak value amplification in a shot-noise-limited interferometer},
Phys. Rev. A {\bf 85}, 062108 (2012).


\bibitem{Bru15}    %{ Bru15,Bru16,How17}
S. Pang and T. A. Brun,
{\it Improving the Precision of Weak Measurements by Postselection Measurement},
Phys. Rev. Lett. {\bf 115}, 120401 (2015).

\bibitem{Bru16}
S. Pang, J. R. G. Alonso, T. A. Brun, and A. N. Jordan,
{\it Protecting weak measurements against systematic errors},
Phys. Rev. A {\bf 94}, 012329 (2016).

\bibitem{How17}
J. Mart\'inez-Rinc\'on, C. A. Mullarkey, G. I. Viza, W. T. Liu, and J. C. Howell,
{\it Ultra sensitive inverse weak-value tilt meter},
Opt. Lett. {\bf 42}, 2479 (2017).



%==新增压制噪声的文献
% Jor17,Pan18,How15,Jor18
\bibitem{Jor17}
J. Sinclair, M. Hallaji, A. M. Steinberg, J. Tollaksen, and A. N. Jordan,
{\it Weak-value amplification and optimal parameter estimation in the presence of correlated noise},
Phys. Rev. A {\bf 96} 052128 (2017).

\bibitem{Pan18}
L. Li, Y. Li, Y.-L. Zhang, S. Yu, C.-Y. Lu, N.-L. Liu, J. Zhang, and J.-W. Pan,
{\it Phase amplification in optical interferometry with weak measurement},
Phys. Rev. A {\bf 97} 033851 (2018).

\bibitem{How15}
G. I. Viza, J. Mart\'{i}nez-Rinc\'{o}n, G. B. Alves, A. N. Jordan, and J. C. Howell,
{\it Experimentally quantifying the advantages of weak-value-based metrology},
Phys. Rev. A {\bf 92} 032127 (2015).

\bibitem{Jor18}
K. Lyons, J. C. Howell, and A. N. Jordan,
{\it Noise suppression in inverse weak value-based phase detection},
Quantum Stud. Math. Found. {\bf 5} 579 (2018).
%

% ======================================================
% \cite{Tana13,FC14,Kne14,Ked12,Jor14,Ren20}
\bibitem{Tana13}
S. Tanaka and N. Yamamoto,
{\it Information amplification via postselection: a parameter-estimation perspective},
Phys. Rev. A {\bf 88}, 042116 (2013).
\bibitem{FC14}
C. Ferrie and J. Combes,
{\it Weak value amplification is suboptimal for estimation and detection},
Phys. Rev. Lett. {\bf 112}, 040406 (2014).
\bibitem{Kne14}  % {Tana13,FC14,Kne14,}
G. C. Knee and E. M. Gauger,
{\it When amplification with weak values fails to suppress technical noise},
Phys. Rev. X {\bf 4}, 011032 (2014).

% \cite{Tana13,FC14,Kne14,Ked12,Jor14,Ren20}
\bibitem{Ked12}
Y. Kedem,
{\it Using technical noise to increase the signal-to-noise ratio of measurements via imaginary weak values},
Phys. Rev. A {\bf 85}, 060102(R) (2012).
\bibitem{Jor14}
A. N. Jordan, J. Martinez-Rincon, and J. C. Howell,
{\it Technical advantages for weak-value amplification: when less is more},
Phys. Rev. X 4, 011031 (2014).
\bibitem{Ren20}
J. Ren, L. Qin, W. Feng, and X. Q. Li,
{\it Weak-value-amplification analysis beyond the Aharonov-Albert-Vaidman limit},
Phys. Rev. A {\bf 102}, 042601 (2020).





%第二部分的参考文献
\bibitem{XQLi22}
Y. Liu, L. Qin, and X.-Q. Li,
{\it Fisher information analysis on weak-value-amplification metrology using optical coherent states},
Phys. Rev. A {\bf 106}, 022619 (2022).

\bibitem{Lui10}
\'A. Rivas and A. Luis,
{\it Precision quantum metrology and nonclassicality in linear and nonlinear detection schemes},
Phys. Rev. Lett. {\bf 105}, 010403 (2010).


%% ============================================================
\bibitem{Ral17}  %{Ral17,Sab17,Sab18}
S. P. Kish and T. C. Ralph,
{\it Quantum limited measurement of space-time curvature
with scaling beyond the conventional Heisenberg limit},
Phys. Rev. A {\bf 96}, 041801(R) (2017).
\bibitem{Sab17}
C. Sab\'in, {\it Quantum detection of wormholes}, Sci. Rep. {\bf 7}, 716 (2017).
\bibitem{Sab18}
C. Sanchidri\'an-Vaca and C. Sab\'in,
{\it Parameter estimation of wormholes beyond the Heisenberg limit},
arXiv:1811.02395v2.     %%; Universe 2016, xx, x; doi:10.3390


%% ============================================================
\bibitem{Kor11}  % {Kor11,Kor16,Li15,Li16}
A. N. Korotkov,
{\it Quantum Bayesian approach to circuit QED measurement},
arXiv:1111.4016; see also chapter 17 in
{\it Les Houches 2011 Session XCVI on Quantum Machines}.
\bibitem{Kor16}
A. N. Korotkov,
{\it Quantum Bayesian approach to circuit QED measurement with moderate bandwidth},
Phys. Rev. A {\bf 94}, 042326 (2016).
\bibitem{Li15}
P. Y. Wang, L. P. Qin, and X.-Q. Li,
{\it Quantum Bayesian rule for weak measurements
of qubits in superconducting circuit QED},
New J. Phys. {\bf 16}, 123047 (2014); {\bf 17}, 059501 (2015).
\bibitem{Li16}
W. Feng, P. F. Liang, L. P. Qin, and X.-Q. Li,
{\it Exact quantum Bayesian rule for qubit measurements in circuit QED},
Sci. Rep. {\bf 6}, 20492 (2016).




%%%  === ZLJ15,Aha15
%\bibitem{ZLJ15}
%L. Zhang, A. Datta, and I. A. Walmsley,
%{\it Precision metrology using weak measurements},
%Phys. Rev. Lett. {\bf 114}, 210801 (2015).
%%%%%%%%%%%%%%%%%%%%%%%%
%%%%%%%%%%%%%%%%%%%%%%%
%%%%%%%%%%%%%%以下文章没有引用


\end{references}
\end{document}